# Physical properties of a quasi-two-dimensional square lattice antiferromagnet $Ba_2FeSi_2O_7$

Tae-Hwan Jang,[1,2] Seung-Hwan Do,[3] Minseong Lee,[4] Vivien S. Zapf,[4] Hui Wu,[5] Craig M. Brown,[5] Andrew D. Christianson,[3] Sang-Wook Cheong,[1,2,6] and Jae-Hoon Park[1,7*]

[1]*MPPHC-CPM, Max Planck POSTECH/Korea Research Initiative, Pohang 37673, Republic of Korea*

[2]*Laboratory for Pohang Emergent Materials and MPPHC-CPM, Pohang 37673, Republic of Korea*

[3]*Materials Science and Technology Division, Oak Ridge National Laboratory, Oak Ridge, Tennessee 37831, USA*

[4]*National High Magnetic Field Laboratory (NHMFL), Los Alamos National Laboratory (LANL), Los Alamos, New Mexico 87545, USA*

[5]*NIST Center for Neutron Research, National Institute of Standards and Technology, Gaithersburg, Maryland 20899, USA*

[6]*Ruters Center for Emergent Materials and Department of Physics and Astronomy, Rutgers University, Piscataway, New Jersey 08854, USA*

[7]*Department of Physics, POSTECH, Pohang 37673, Republic of Korea*

**ABSTRACT**

We report the magnetization ($\chi$, $M$), specific heat ($C_P$), and neutron powder diffraction results on a quasi-two-dimensional $S = 2$ square lattice antiferromagnet $Ba_2FeSi_2O_7$ consisting of $FeO_4$ tetragons with a large compressive distortion (27%). Despite of the quasi-two-dimensional lattice structure, both $\chi$ and $C_P$ present three dimensional magnetic long-range order below the Néel temperature $T_N = 5.2$ K. Neutron diffraction data shows a collinear $\boldsymbol{Q}_m = (1,0,0.5)$ antiferromagnetic (AFM) structure with the in-plane ordered magnetic moment suppressed by 26% below $T_N$. Both the AFM structure and the suppressed moments are well explained by the Monte Carlo simulation with a large single-ion *ab*-plane anisotropy $D = 1.4$ meV and a rather small in-plane Heisenberg exchange $J_{intra} = 0.15$ meV. The characteristic two dimensional spin fluctuations can be recognized in the magnetic entropy release and diffuse scattering above $T_N$. This new quasi-2D magnetic system also displays unusual non-monotonic dependence of the $T_N$ as a function of magnetic field $H$.

**I. INTRODUCTION**

Two-dimensional (2D) Heisenberg antiferromagnets have been intensively studied both in theory and in experiment to exploring exotic low-dimensional magnetic behaviors. Mermin-Wagner theorem states that no long range magnetic order can be stablized at finite temperature in the 2D Heisenberg magnetic system due to strong spin fluctuations [1]. However, lattice topology and strong magnetic anisotropy are predicted to be able to realize a 2D Heisenberg antiferromagnet [2] as in 2D-Ising and 2D-XY spin systems [3-7] under an external magnetic field. On the other hand, three-dimensional long-range magnetic ordering have often been observed in many layered magnetic materials [8-10] because of the quasi-2D nature with minimal but non-vanishing interlayer magnetic coupling [11, 12].

The Melilite family of compounds $A_2MB_2O_7$ (A = Ca, Sr, Ba, $M$ = divalent 3*d* transition metals, B = Si, Ge) are interesting examples of quasi-2D square lattice Heisenberg antiferromagnetic systems.

The *d-p* metal-ligand hybridization induces distinct magnetoelectricity [13], directional dichroism involving spin wave/optical excitations [14], magnetochiral effect [15], and longitudinal magnon mode associated with electromagnon [16, 17]. Most studies have been performed on Meililite compounds with half-integer spin quantum numbers, $M$ = Mn (5/2), Co (3/2), Cu(1/2) in last decades. Meanwhile, the studies on the compounds with an integer spin number such as $M$ = Ni$^{2+}$ ($S$ = 1) or Fe$^{2+}$ ($S$ = 2) have rarely been carried out due to lack of reliable crystal quality samples, and only a few ones have been reported recently; a theoretical work on Jahn-Teller distortion driven ferroelectricity in Ba$_2$NiGe$_2$O$_7$ [18] and a THz experimental one on spin-orbital excitations in Sr$_2$FeSi$_2$O$_7$ [19]. Especially, the Fe ($S$ = 2) based compounds present strongly compressed FeO$_4$ tetrahedrons along the *c*-axis suggesting intriguing magnetic properties governed by a non-trivial magnetic gap [19-22].

Here we present physical properties of a new quasi-2D square lattice integer spin ($S$ = 2) antiferromagnet Ba$_2$FeSi$_2$O$_7$. As shown in Fig. 1(a), this compound is crystallized in the $P\bar{4}2_1$m tetragonal melilite-type structure with the lattice constants $a$ = 8.3261 Å, $c$ = 5.3401 Å at room temperature [23]. The system is composed of FeO$_4$ tetrahedra connected via SiO$_4$ polyhedra and the FeSi$_2$O$_7$ layers are separated by Ba layer, forming a quasi 2D square-lattice structure. The magnetic coupling is dominated by the intra-layer Heisenberg interaction ($J_{\text{intra}}$) through the neighboring Fe$^{2+}$-O$^{2-}$-O$^{2-}$-Fe$^{2+}$ exchange path and the layered structure contributes minimal inter-plane interaction ($J_{\text{inter}}$), resulting in a quasi-2D magnetic system. Noticeably, the FeO$_4$ tetrahedron is compressed as large as 27% along the *c*-axis with respect to the perfect tetrahedron. Such a large compression splits both the triplet t$_{2g}$ and doublet e$_g$ orbital states and make the unquenched orbital angular momentum considerable, which is responsible for noticeable single-ion anisotropy ($D$) [20, 21]. Considering that $D \sim$ 1.1 meV was estimated in Ba$_2$CoGe$_2$O$_7$ with 13% compression of the CoO$_4$ tetrahedron [16, 24, 25], the $D$ value is expected to be somewhat enhanced in Ba$_2$FeSi$_2$O$_7$ with 27% compression of the FeO$_4$ tetrahedron.

In this article, we report the physical properties of a new quasi-2D magnetic material Ba$_2$FeSi$_2$O$_7$

using magnetization, specific heat, and neutron powder diffraction measurements. The results manifest an AFM ordering below the Néel temperature ($T_N$ = 5.2 K) with large easy plane magnetic anisotropy. From Monte Carlo simulation, we estimate $J_{intra}/D \sim 0.1$. Specific heat measurement reveals a Schottky anomaly arising from thermal population of the low-lying excited spin-orbital states. Neutron diffraction measurements reveals that the short-range spin correlations appears below 20 K and that antiferromagnetic structure is characterized by a staggered magnetic moment of 2.95 $\mu_B$, which is considerably smaller than the moment (4 $\mu_B$) expected from $S$ = 2. The field dependent measurements exhibit a unusual non-monotonic behavior of $T_N(H)$ as a function of the $H$-field, indicating that a quasi-2D square lattice magnet of $Ba_2FeSi_2O_7$ is a novel easy-planar integer spin system.

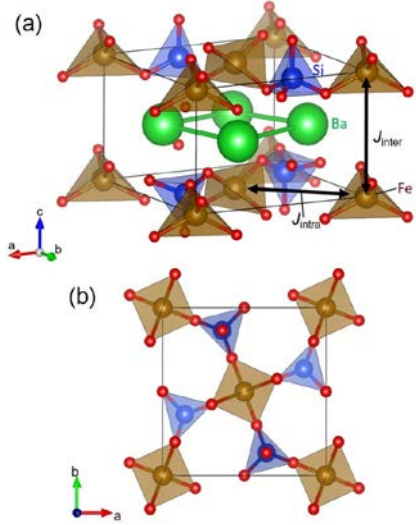

FIG. 1. (a) The crystal structure of $Ba_2FeSi_2O_7$ obtained from the Rietveld refinement of the structural model against the neutron diffraction data. Thick black arrows indicate nearest neighbor Heisenberg exchange interaction within $ab$-plane ($J_{intra}$) and between adjacent inter-planes ($J_{inter}$). (b) A single layer of $FeSi_2O_7$ projected in $ab$-plane. The exchange path between $Fe^{2+}$ spins is through two oxygen ligands.

## II. METHODS

To obtain single crystals of $Ba_2FeSi_2O_7$, we prepared a polycrystalline of $Ba_2FeSi_2O_7$ as a precursor using the solid-state reaction. Stoichiometric mixture of $BaCO_3$, $Fe_2O_3$, and $SiO_2$ were thoroughly ground, pelletized, and heated at 1050°C with intermediate sintering. X-ray and neutron powder diffraction measurements on the polycrystalline samples identified dominant phase of $Ba_2FeSi_2O_7$ (96.5%) with minor $Ba_2SiO_4$ (2.6%) and $SiO_2$ (0.9%) (see Fig. 6). The polycrystalline samples were prepared as feed rods, and single crystal of $Ba_2FeSi_2O_7$ was grown using floating zone melting method under reducing gas atmosphere. Growth direction was perpendicular to *c*-axis and size of the as-grown crystals was about 8 mm in diameter and 60 mm in length. The powder XRD pattern on crushed crystals presents a single phase of $Ba_2FeSi_2O_7$, as described in Appendix A.

Temperature (*T*) and magnetic field (*H*) dependence of *dc*-magnetization and specific heat measurements on a $Ba_2FeSi_2O_7$ crystal were performed by using a vibrating sample magnetometry (VSM) option and a standard calorimetric relaxation technique equipped in a physical property measurement system (PPMS) of Quantum Design DynaCool-9 T. The magnetization results were compared with classical Monte Carlo simulations to estimate the energy scale of the interactions in $Ba_2FeSi_2O_7$. For the calculation, a square lattice with size of 16 × 16 × 6 with periodic boundary conditions was employed.

Neutron powder diffraction measurements were carried out by using the BT-1 High-Resolution Powder Diffractometer (HRPD) at NIST Center for Neutron Research (NCNR), USA. A 2.9 g of polycrystalline sample was loaded into a vanadium can and cooled using flow type cryostat. Constant wavelength $\lambda$ = 2.0772 Å of neutron beam was collimated using Ge (311)-60° monochromator. Diffraction data were collected at temperatures, 1.7, 3, 8, 10, 20, and 30 K. The refinement was carried out by the Rietveld methods using the FULLPROF program [26], and software SARA*h* was used for representational analysis to determine symmetry-allowed magnetic structures [27].

## III. EXPERIMENTAL RESULTS

## A. Magnetic properties

Figure 2(a) shows the temperature dependence of magnetic susceptibility ($\chi = M/H$) for a $Ba_2FeSi_2O_7$ single crystal with magnetic field parallel ($H\|ab$) and perpendicular ($H\|c$) to the $ab$-plane. The magnetic susceptibility exhibits strongly anisotropic easy-planer spin behavior over a broad temperature range. The $ab$-plane is magnetic easy-plane and $c$-axis is hard axis. At low temperatures, $\chi(T)$ for both field directions exhibit peaks around $T \sim 8$ K, indicating the onset of the short-range magnetic order with the 2D spin fluctuations. The AFM long-range ordering temperature is determined to be $T_N = 5.2$ K from the first derivative of the in-plane magnetic susceptibility ($d\chi/dT$), which exhibits the sharp anomaly (see the inset in Fig. 2(a)).

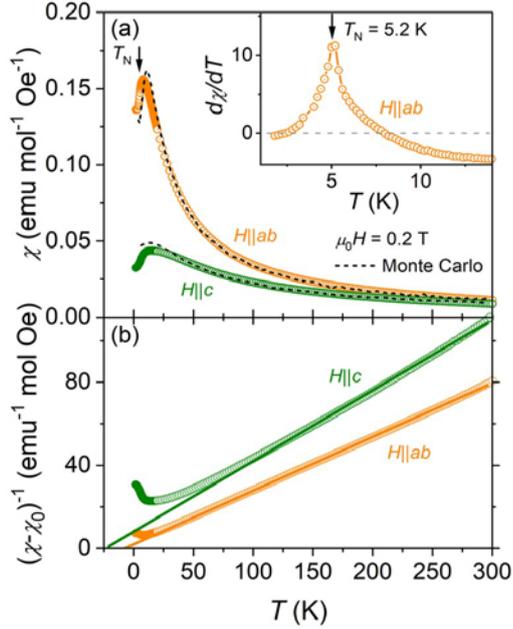

FIG. 2. (a) Temperature ($T$) dependence of dc magnetic susceptibility ($\chi = M/H$) with applied magnetic fields $\mu_0 H = 0.2$ T along $H\|ab$ (orange symbol) and $H\|c$ (olive symbol). The dashed lines are the results of classical Monte Carlo simulations with $J_{intra} = 0.15$ meV, $J_{inter} = 0.0025$ meV, $D = 1.4$ meV with field of 0.2 T. Inset: The first derivative of the magnetic susceptibility ($d\chi/dT$) as a function of temperature. Vertical arrows in (a) denote the magnetic transition temperature at $T_N = 5.2$ K. (b) Inverse magnetic susceptibility with magnetic field along $H\|ab$ (orange symbol) and $H\|c$ (olive symbol). Solid lines are Curie-Weiss fit to the data from 100 K to 300 K.

The inverse magnetic susceptibilities in Fig. 2(b) show linear behaviors above 100 K, in which the susceptibility can be described by the Curie-Weiss formula, $\chi(T) = \chi_0 + C/(T - \Theta_{CW})$, where $C$ and $\Theta_{CW}$ are the Curie constant and the Curie-Weiss temperature, respectively. $\chi_0$ is the diamagnetic background susceptibility. The fitting to the experimental magnetic susceptibility in the temperature range from 100 K to 300 K gives an effective magnetic moment $\mu_{eff}$ [$ab$] = 5.56(1) $\mu_B$, $\mu_{eff}$ [$c$] = 4.84(1) $\mu_B$ and Curie-Weiss temperatures $\Theta_{CW}$ [$ab$] = -7.4(2) K, $\Theta_{CW}$ [$c$] = -23.7(2) K, respectively. The out-of-plane moment is comparable to the spin only value of $S = 2$ ($\mu_{eff} \sim 4.9$ $\mu_B$ for $g = 2$) while the in-plane one $\mu_{eff}$ [$ab$] is considerably larger than the value. It implies that an unquenched angular momentum is present and makes anisotropic contribution to the magnetic moment [21], in consistent with the observed anisotropic behavior of $\chi$ even up to room temperature. The obtained Curie-Weiss temperature larger than $T_N$ ($\Theta_{CW}$ [$c$] > 4$T_N$) is attributed to the spin fluctuation involving spin-spin interaction with a strong 2D character.

Figure 3 (a) presents isothermal magnetization $M(H)$ as a function of magnetic field $H$ up to 9 T for $H\|ab$ (110) and $H\|c$ (001) at $T = 1.8$ K. $M(H)$ shows large anisotropy for $H\|ab$ and $H\|c$ reflecting the strong easy-planer spin, but both $M_{ab}(H) \equiv M(H\|ab)$ and $M_c(H) \equiv M(H\|c)$ almost linearly increase as $H$ increases. Interestingly, the slope in $M_{ab}(H)$ changes considerably around $\mu_0 H \sim 0.3$ T ($\mu_0 H_{ab1}$) and $\sim 7.4$ T ($\mu_0 H_{ab2}$). As shown in Fig. 3(b), these anomalies become more noticeable in its derivative $dM_{ab}/dH$ while those disappear at 6.5 K (> $T_N$), indicating that there exist two field-induced transitions below $T_N$. On the other hand, $M_c$ monotonically increases with $H$-field up to 9 T without any noticeable anomaly representing the field induced transition.

The low field transition at $H_{ab1}$ can be attributed to a spin-flop-like transition aligning two AFM domains. At $H = 0$ field, there exist equally populated two AFM domains; the AFM ordered spins along in-plane easy-axis (110) in one and (1-10) in the other (AFM-I). At $H$ increases across $H_{ab1}$, the spin axes of both domains align to be perpendicular to the $H$-direction in the $ab$-plane (AFM-II). A similar transition was also observed in Ba$_2$CoGe$_2$O$_7$ [25]. We note that $H_{ab1}$ shows a minimal azimuthal angle dependence in the plane, indicating that the in-plane magnetic anisotropy is minimal. $H_{ab1}$ exhibits almost no temperature dependence below $T_N$ (not shown here) and disappears above $T_N$. On the other hand, the high field transition enhances $M_{ab}$ across $H_{ab2}$, and the enhanced magnetic moment $\Delta M_{ab}(T) = M_{ab}(T) - M_{ab}(6.5\ K)$ is estimated to be $\sim 0.15\ \mu_B/Fe^{2+}$ at $T = 1.8\ K$. To trace the anomalies, we measured $M_{ab}(H)$ at different temperatures below $T_N$. The inset shows $dM_{ab}/dH$ as a function of $H$ at various temperatures below $T_N$. As temperature increases, the $dM/dH$ peak feature becomes weaker and $H_{ab2}$ shifts to lower fields. The peak disappears above $T_N$, indicating this transition is also relevant to the AFM phase.

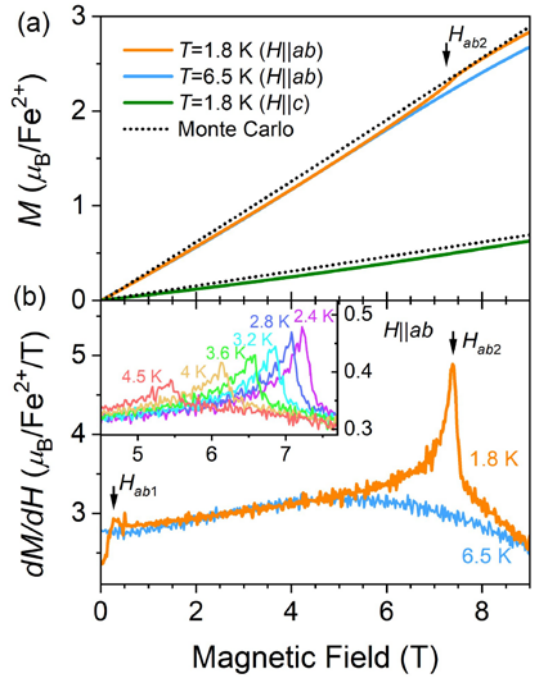

FIG. 3. (a) Magnetic field ($H$) dependence of magnetization ($M$) curve along $H\|ab$ (orange and blue symbols for $T = 1.8\ K$ and 6.5 K, respectively) and $H\|c$ (olive symbol, $T = 1.8\ K$). The dotted lines are the results of classical Monte Carlo simulations with $J_{intra} = 0.15$ meV, $J_{inter} = 0.0025$ meV and $D = 1.4$ meV (for detailed information about calculation, see the section IV.). (b) First derivative of magnetization curve ($dM/dH$) as a function of magnetic field along $H\|ab$ measured at 1.8 K (orange symbol) and 6.5 K (blue symbol). Two vertical arrows indicate positions of two critical magnetic fields ($\mu_0 H_{ab1} \sim 0.3$ T, $\mu_0 H_{ab2} \sim 7.4$ T) showing $H$ induced weak and sharp peak of $dM/dH$ at $T = 1.8\ K$, respectively. The inset in (b) shows

*dM/dH* as a function of magnetic field along *H*||*ab* measured at various temperatures below $T_N$.

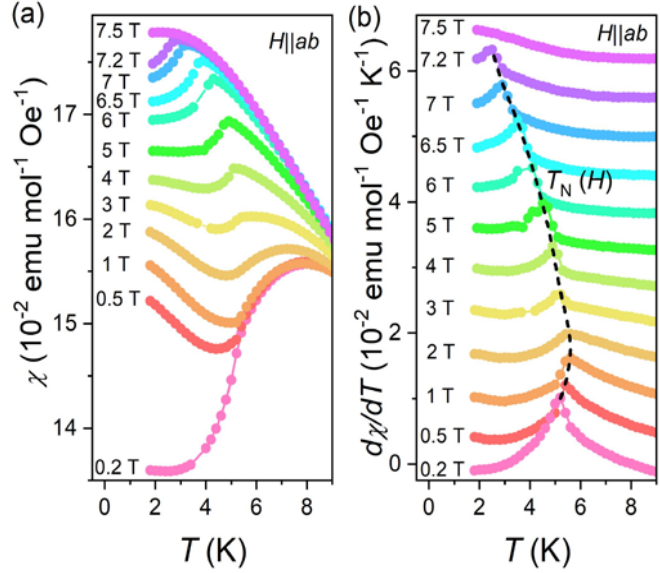

FIG. 4. (a) Temperature (*T*) dependence of dc magnetic susceptibility ($\chi = M/H$) and (b) first derivative of magnetic susceptibility ($d\chi/dT$) as a function of temperature (*T*) for applied magnetic fields along *H*||*ab*. In (b), the dotted guide line is use to indicate $T_N$ as a function of *H* obtained from the peak position in $d\chi/dT$. For clarity, each $d\chi/dT$ curve is vertically shifted by 0.006 emu mol$^{-1}$ Oe$^{-1}$ K$^{-1}$.

Figures 4 (a) shows $\chi(T)$ as a function of temperature *T* measured at various *H*||*ab* fields up to $\mu_0 H$ = 7.5 T. $\chi(T)$ below $T_N$ suddenly changes between 0.2 T and 0.5 T due to the spin-flop-like transition across $\mu_0 H_{ab1}$ ~ 0.3 T observed $M_{ab}$ ($dM_{ab}/dH$). As presented in Fig. 4(b), the derivatives $d\chi/dT$ clearly exhibit peak features representing the AFM transition up to $\mu_0 H$ = 7.2 T ($< \mu_0 H_{ab2}$) and enable us to determine $T_N(H)$ at a given *H*||*ab* field. Interestingly, $T_N(H)$ shows a non-monotonic field dependent behavior. $T_N(H)$ slightly increases, as *H* increases up to $\mu_0 H$ ~ 2 T, and then it decreases for further increasing *H* up to 7.2 T. At $\mu_0 H$ = 7.5 T ($> \mu_0 H_{ab2}$), the $d\chi/dT$ peak feature becomes completely suppressed with saturation in $\chi(T)$.

## B. Specific heat

Figure 5(a) shows total specific heat ($C_P$) of $Ba_2FeSi_2O_7$ at $H = 0$. Lattice contribution ($C_L$) was estimated using Debye-Einstein model, where $C_L(T)$ is defined by [28, 29],

$$C_L(T) = m \left[ 9Rx_D^{-3} \int_0^{x_D} \frac{x^4 e^x}{(e^x - 1)^2} \right] + \sum_{i=1}^{s-1} n_i \left[ 3R \frac{x_{E_i}^2 e^{x_{E_i}}}{(e^{x_{E_i}} - 1)^2} \right] \qquad (1)$$

The first term represents the Debye specific heat for the acoustic phonon modes and the second term represent the Einstein specific heat for optic phonon modes. $x_D$ and $x_{Ei}$ are defined as $x_D = \Theta_D/T$ and $x_{Ei} = \Theta_{Ei}/T$ where $\Theta_D$ and $\Theta_{Ei}$ are the Debye temperature and the Einstein temperatures, respectively. The constant $m$ and $n_i$ are the number of degrees of freedom for each contribution and $R$ is the molar gas constant. We performed fitting Eq. (1) to the experimental data in the temperature range from 70 K to 250 K, which yields $\Theta_D \sim 237$ K ($m = 4.8$), $\Theta_{E1} \sim 554$ K ($n_1 = 4.3$) and $\Theta_{E2} \sim 1345$ K ($n_2 = 2.9$) with $m + n_1 + n_2 = 12$ (total number of atoms in the formula unit). Based on these fitting parameters, the extracted $C_L$ is displayed in Fig. 5(a). Magnetic specific heat ($C_M$) shown in Fig. 5(b) was obtained by subtracting the lattice contribution from the total specific heat, i.e. $C_M = C_P - C_L$. $C_M$ displays a sharp $\lambda$-anomaly at $T_N = 5.2$ K, which coincides with $T_N$ determined from the magnetic susceptibility. Above $T_N$, $C_M$ exhibits a broad peak around $T_{SO} \sim 8$ K, which represents the short-range ordering in the quasi 2D spin system where the long-range order is suppressed by the low-dimensionality [2].

The magnetic entropy, $\Delta S_M(T)$, was calculated by using $\Delta S_M(T) = \int_0^T \Delta C_M(T)/T dT$. $\Delta S_M$ at $T = 50$ K is obtained to be 12.74 J mol$^{-1}$ K$^{-1}$ that corresponds to 95% of $R\ln(2S + 1) = R\ln5$, the total entropy of $S = 2$. We note that only about 20% of the total entropy is released at $T_N$ and additional entropy involving the short range order is released by above the transition temperature ($T_{SO} \sim 8$ K). Interestingly, the entropy $R\ln(3) \sim 9.13$ J mol$^{-1}$ K$^{-1}$ corresponding to the degree of freedom for $S = 1$ releases up to around 18 K, where the short-range ordering peak diminishes. Above this temperature, a Schottky-like broad peak is visible in $C_M$ around $T_{Broad} \sim 23$ K and the entropy gradually releases remaining the spin degree of freedom for $S = 2$ until even above 50 K.

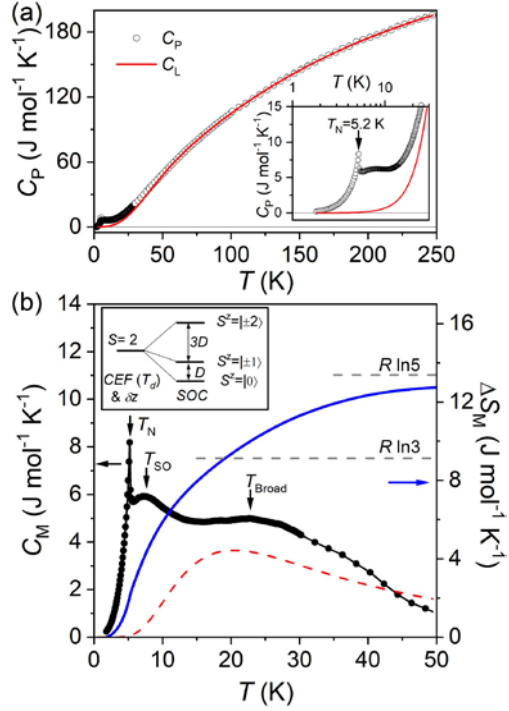

FIG. 5. (a) Total specific heat ($C_P$). Open circle and red line display the measured total specific heat ($C_P$) and calculated lattice contribution of specific heat ($C_L$), respectively. Inset magnifies the specific heat below 30 K in semi-logarithmic scale. The vertical arrow indicates the magnetic transition temperature ($T_N = 5.2$ K). (b) Magnetic specific heat ($C_M$) and magnetic entropy gain ($\Delta S_M$) as a function of temperature ($T$). Above $T_N$, $C_M$ shows two broad peaks centered at $T_{SO} \sim 8$ K and $T_{Broad} \sim 23$ K, associated with short-range spin correlations and a Schottky anomaly from the excitation between the levels given by single-ion state, respectively. Two gray horizontal dashed lines show $R\ln(2S + 1)$ with spin state $S = 1$ ($R\ln3$) and $S = 2$ ($R\ln5$), respectively. Inset represents the level structure of the lowest-energy $d_{z^2}$ orbital for $Fe^{2+}$ ion in the tetrahedral crystal field ligand ($T_d$) with a tetragonal compression ($\delta z$) and further splitting energy level of the spin states ($S^z$) by the 2nd order spin-orbit coupling (SOC) [19, 20]. The red dashed curve indicates the calculated Schottky anomaly for the transition between $S^z = |\pm 2\rangle$ and $S^z = |\pm 1\rangle$ states with gap, $\Delta = 3D = 3*1.4$ meV $= 4.2$ meV where $D$ is referred from the Monte Carlo calculations (see the text).

## C. Powder neutron diffraction

To study the AFM spin structure below $T_N$, we have carried out zero field ($H = 0$) neutron powder diffraction (NPD) measurements on $Ba_2FeSi_2O_7$. Figure 6 shows the NPD patterns at 30 K ($> T_N$) and 1.7 K ($< T_N$). The crystal and magnetic structures were determined from the Rietveld refinement fitting by using FULLPROF [26]. The refined crystallographic parameters are tabulated in Table I ($T = 30$ K) and II ($T = 1.7$ K). Both the $T = 30$ K and $T = 1.7$ K diffraction patterns for the nuclear Bragg peaks are well described by the tetragonal space group, $P\bar{4}2_1m$ (SG: 113), and the Bragg peak profiles exhibit only small variations across $T_N$, evidencing that the AFM transition does not accompany any considerable structural transition. Comparing low $Q$-region (0.5 Å$^{-1}$ ≤ $Q$ ≤ 2.0 Å$^{-1}$) neutron diffraction patterns at $T = 1.7$ K and 30 K as shown in the inset, we identify the magnetic Bragg reflections at $Q$ = (1,0,1/2) and (2,1,1/2) below $T_N$, indicating a characteristic vector of $Q_m$ = (1,0,1/2).

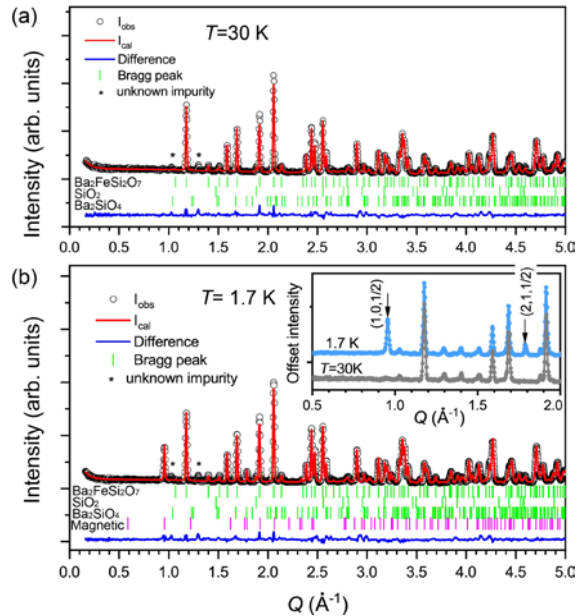

FIG. 6. Neutron powder diffraction patterns for $Ba_2FeSi_2O_7$ at (a) $T = 30$ K ($> T_N$) and (b) 1.7 K ($< T_N$). Open circles and the red solid line represent the experimental data and the Rietveld refinement fitting line, respectively. At both temperatures, Bragg peaks from $SiO_2$ and $Ba_2SiO_4$ (non-magnetic secondary phases) are visible in the sample, and the Rietveld refinement quantifies the phase fractions of 0.9% and 2.6%, respectively. In (b), the structural and magnetic Bragg reflections are presented by upper (green) and low (violet) ticks, respectively. The inset shows an expanded view of the low-$Q$ region data and miller indexed

TABLE I. Crystallographic parameters with space group $P\bar{4}2_1m$ (SG:113) from Rietveld refinements on the diffraction data at $T = 30$ K. Lattice constants $a = b = 8.3193(8)$ Å, $c = 5.3348(5)$ Å, and $\alpha = \beta = \gamma = 90°$. $R_{wp} = 6.75\%$.

| atom | site | x | y | z | B |
|---|---|---|---|---|---|
| Ba | 4e | 0.1648(3) | 0.6648(3) | 0.5090(6) | 0.05(12) |
| Fe | 2a | 0 | 0 | 0 | 0.05(4) |
| Si | 4e | 0.3627(3) | 0.8627(3) | 0.9610(7) | 0.12(12) |
| O1 | 2c | 0 | 0.5 | 0.1371(8) | 0.36(8) |
| O2 | 8f | 0.3649(3) | 0.8649(3) | 0.2627(5) | 0.17(6) |
| O3 | 4e | 0.0764(3) | 0.1990(2) | 0.1712(4) | 0.15(5) |

TABLE II. Crystallographic parameters with space group $P\bar{4}2_1m$ (SG:113) from Rietveld refinements on the diffraction data at $T = 1.7$ K. Lattice constants $a = b = 8.3194(2)$ Å, $c = 5.3336(5)$ Å, and $\alpha = \beta = \gamma = 90°$. $R_{wp} = 7.14\%$.

| atom | site | x | y | z | B |
|---|---|---|---|---|---|
| Ba | 4e | 0.1644(3) | 0.6644(3) | 0.5098(7) | 0.08(10) |
| Fe | 2a | 0 | 0 | 0 | 0.20(5) |
| Si | 4e | 0.3645(3) | 0.8645(3) | 0.9609(7) | 0.11(8) |
| O1 | 2c | 0 | 0.5 | 0.1383(8) | 0.54(9) |
| O2 | 8f | 0.3651(3) | 0.8651(2) | 0.2642(5) | 0.07(6) |
| O3 | 4e | 0.0769(3) | 0.1984(2) | 0.1694(5) | 0.32(5) |

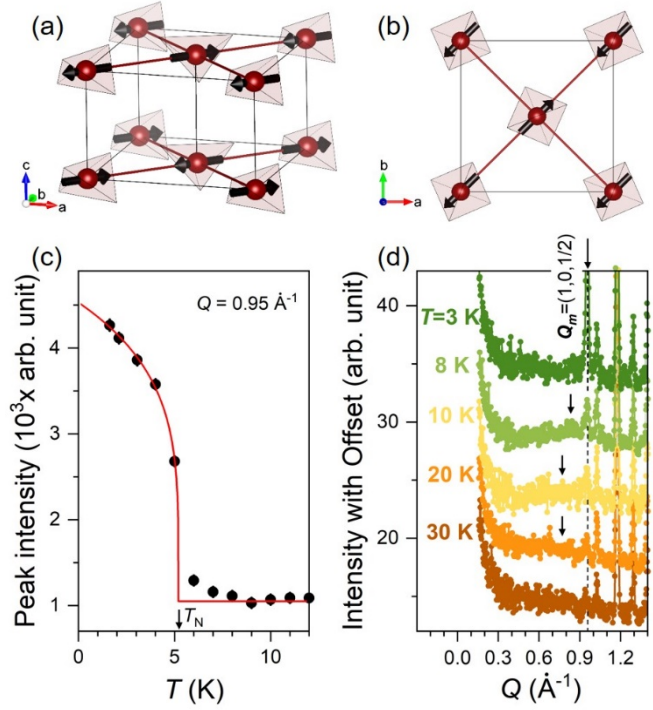

FIG. 7. (a), (b) The magnetic structure of $Ba_2FeSi_2O_7$. The structure has collinear spin alignment of Fe spins with $\boldsymbol{Q_m}$ = (1,0,1/2) (= 0.95 Å$^{-1}$). (c) Magnetic peak intensity at $Q$ = 0.95 Å$^{-1}$ as a function of temperature (black closed circles). The red solid line is a guide to eye, and $T_N$ = 5.2 K is indicated by a vertical arrow. Near constant intensity above $T_N$ reflects the structural contribution at $Q$. (d) Neutron powder diffraction patterns at temperatures as indicated in the figure.

Representation analysis was used to determine symmetry-allowed magnetic structures. Irreducible representations $\Gamma_{mag} = 1\Gamma^1_1 + 1\Gamma^1_2 + 2\Gamma^2_5$ are compatible with the $P\bar{4}2_1m$ symmetry with two Fe sites at (0,0,0) and (1/2,1/2,0). The basis vectors of $2\Gamma^2_5$ describe all of magnetic Bragg peaks with a collinear antiferromagnetic spin structure as depicted in Fig. 7(a) and (b). The in-plane collinear AFM spin alignment indicates that the nearest neighbor spin-spin interaction is governed by the Heisenberg $J_{intra}$ through $Fe^{2+}$-$O^{2-}$-$O^{2-}$-$Fe^{2+}$ exchange path in the plane (see Fig. 1). The ordered magnetic moment of $Fe^{2+}$ ion is determined to be 2.95 $\mu_B$, which is only 74% of the full moment of $Fe^{2+}$ spin ($S$ = 2).

Figure 7(c) shows evolution of the magnetic peak intensity at $Q_m$ = (1,0,1/2) ($Q$ = 0.95 Å$^{-1}$) with temperature. Figure 7(d) presents NPD $Q$-scans at different temperatures from 3 K to 30 K. The sharp and intense magnetic Bragg peak, which is present at $Q = Q_m$ = 0.95 Å$^{-1}$ in the 3 K scan, mostly diminishes at 8 K and becomes negligible at higher temperatures. Besides the remnant of the sharp magnetic peak, an additional broad peak feature is observable around $Q$ = 0.8 Å$^{-1}$ (marked with a black arrow) in the 8 K scan. This feature gradually fades out and shifts to low $Q$ upon heating, and then finally disappears at 30 K, well above $T_N$. This $Q$-dependent diffusive scattering is attributed to short range spin-spin correlations, which were also observed in the magnetic specific heat $C_M$ ($T$) as a broad peak feature around $T_{SO}$ ~ 8 K (see Fig. 5). Presence of the diffusive scattering feature reflects strong spin fluctuations in the low dimensional quasi-2D magnetic system.

## IV. DISSCUSSION

We observe multiple magnetic transitions with temperature and in-plane magnetic fields ($H \| ab$) in the magnetization and specific heat measurements. Those transitions can be summarized with a phase diagram in an $H$-$T$ space as shown in Fig. 8. The phase boundaries are defined by the peak positions determined from $d\chi/dT$, $dM/dH$, and $C_M$. At a zero field, the system is in the AFM-I phase with two types of AFM domains below $T_N$, and transits to the paramagnetic phase (PM) upon heating across $T_N$. On the other hand, as $H$ increases across $\mu_0 H_{ab1}$ ~ 0.3 T well below $T_N$, the AFM-I phase transits to the AFM-II phase with a single type of AFM domains. The AFM ordered spins, which lie to be nearly perpendicular to the $H$-direction, slightly cant toward the $H$-direction and result in a finite $M$, i.e. a composition of AFM and ferromagnetic (FM) components (field induced canted AFM). As the $H$-field further increases, the AFM component decreases and finally disappears. The AFM-II phase transits to the spin polarized (SP) phase across $H_{ab2}$ with a certain gain of $\Delta M$. $\mu_0 H_{ab2}$ ~ 7.4 T determined from $M(H,T)$ at 1.8 K decreases as $T$ increases (see Fig. 3). $H_{ab2}(T)$ nearly coincides with $T_N(H)$ from $\chi(H,T)$ (see Fig. 4) up to $T$ ~ 4 K. Upon further heating, $H_{ab2}(T)$ somewhat deviates from $T_N(H)$ and finally disappears at $T > \sim 5$ K (or $\mu_0 H < \sim 4$ T), implying that the SP phase crosses over to the PM phase.

We note that $T_N = 5.2$ K at $H = 0$ increases up to 2 T and then decreases above 2 T as $H$ increases. This non-monotonic behavior of $T_N(H)$ was also observed in other quasi-two dimensional spin systems with very weak inter-layer exchange coupling ($J_{\text{inter}}$) [9, 10]. At a low magnetic field, the $S^z$ spin fluctuation becomes suppressed and the spin correlation within the $ab$-plane becomes effectively enhanced to increases $T_N$. At a high field, the spin canting effect prevails to reduce $T_N$ as usual. Appearance of the non-monotonic $T_N(H)$ manifests that $Ba_2FeSi_2O_7$ is a spin system with strong 2D character. It is also consistent with remarkable short-range spin correlation above $T_N$ observed in specific heat and neutron diffraction results.

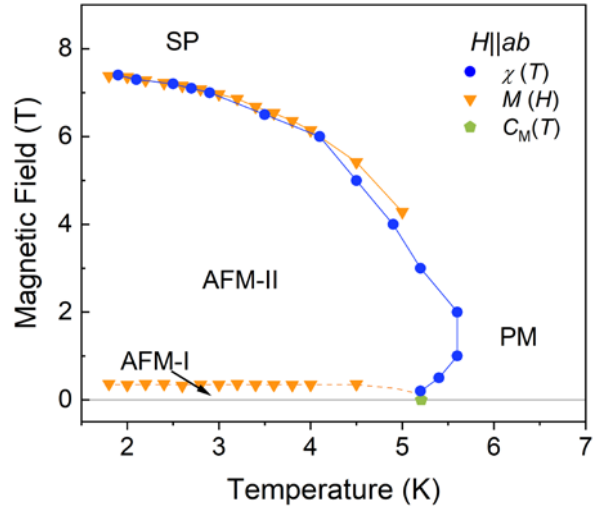

FIG. 8. Magnetic phase diagram of the $Ba_2FeSi_2O_7$ with applied magnetic field $H \| ab$. Blue and green symbols present $T_N$ determined from the magnetic susceptibility and specific heat measurements, respectively. Orange symbols represent critical magnetic fields ($H_{ab1}$, $H_{ab2}$) determined from the magnetization measurements. AFM-I, AFM-II, SP, and PM denote antiferromagnetic (two types of AFM domains), field induced canted antiferromagnetic, spin-polarized, and paramagnetic phases, respectively.

To quantify energy scales of the exchange interactions and single-ion anisotropy, we performed the Monte Carlo simulations to calculate the magnetic properties. The calculated $\chi(T)$ and $M(H)$ are compared with the corresponding experimental ones in Fig. 2 and Fig. 3, respectively. For the simulation, we constructed a simple spin Hamiltonian consisting only with Heisenberg exchange interactions, a single-ion anisotropy, and a Zeeman term for $S = 2$ as follows;

$$\mathcal{H} = J_{\text{intra}} \sum_{<ij>_1} \mathbf{S}_i \cdot \mathbf{S}_j + J_{\text{inter}} \sum_{<ij>_2} \mathbf{S}_i \cdot \mathbf{S}_j + D \sum_i (S_i^z)^2 - \mu_B \sum_i \mathbf{S}_i \cdot \mathbf{g} \cdot \mathbf{B} ,$$

where $<i, j>_1$ and $<i, j>_2$ denote the in-plane and inter-plan nearest neighbors, respectively. The direction of $z$ is parallel to the $c$-axis (see Fig. 1(a)). Although it is not possible to uniquely determine values of the exchange parameters, we could quantify $J_{\text{intra}} = 0.15$ meV, $J_{\text{inter}} = J_{\text{intra}}/60$, and $D = 1.4$ meV, $g_{ab} = 2.6$ and $g_c = 2.3$, which fairly well reproduce $T_N$, high temperature $\chi(T)$ above 50 K (Fig. 2(a)), and the magnetic anisotropy $M(H)$ (Fig. 3(a)). $\chi(T)$ below 50 K deviates from the Curie Weiss formula. We ascribe this deviation to thermal depopulation of the high energy spin states split by the strong single-ion anisotropy, which our classical Monte Carlo simulations do not account for.

Together with tetragonal compression of FeO$_4$ tetrahedrons along the $z$-direction in Ba$_2$FeSi$_2$O$_7$, the spin-orbit coupling (SOC) splits the $S = 2$ state with $(2S +1)$-fold degeneracy into one singlet ground state ($S^z = 0$) and two doublet exicited states ($S^z = | \pm 1>$ and $S^z = | \pm 2>$) with finite gaps of $D$ and $3D$, respectively (see inset in Fig. 5(b)) [19-21]. Hence these low-lying ground/excited spin states are governed by thermal populations in the temperature range of $4D$ (5.6 meV $\sim$ 70 K) energy scale. The residual broad peak around 23 K in the magnetic specific heat is considered to be associated with the thermal populations of $S^z = | \pm 1>$ and $S^z = | \pm 2>$ states. The Schottky anomaly for the gap $\Delta = 3D$ with $D = 1.4$ meV from the Monte Carlo simulation (red dashed line in Fig. 5(b)) reproduces the peak position and width of the observed broad peak, which is consistent with $D$ value obtained from the recent inelastic neutron scattering analysis [22]. Thermal populations of the two excited states ($S^z = | \pm 1>$ and $S^z = | \pm 2>$) were also similarly observed in the THz absorption data of a sister compound Sr$_2$FeSi$_2$O$_7$ (denoted by $\beta$-mode absorption) [19]. It is worth to note that magnetic susceptibility along $c$-direction deviates from the Curie-Weiss formula below 70 K, which is consistent with the temperature of the onset of the Schottky anomaly peak. The deviations in $\chi_c$ and the Schottky peak evidence the presence of a single-ion anisotropy in $Ba_2FeSi_2O_7$.

## V. CONCLUSION

In summary, we have studied the effects of the large single ion anisotropy ($D$) on the physical properties in the new $S = 2$ quasi-2D square lattice antiferromagnet $Ba_2FeSi_2O_7$ with $M$, $\chi$, $C_M$, and NPD measurements. The spin states gapped with $D$ and their thermal populations are responsible for the remarkable 2D spin fluctuations such as Schottky anomaly and short-range magnetic ordering with strong release of the magnetic entropy gain. On the other hand, below $T_N = 5.2$ K, $M$ and $\chi$ exhibits large easy plane type anisotropy behaviors, and the NPD data yields a significantly reduced magnetic ordered moment. For integer spin systems with a large easy-plane type single-ion anisotropy, the paramagnetic ground state with the finite gaps is possibly realized by the local $S^z = 0$ state favoring quantum disordered state, $Ba_2FeSi_2O_7$ is suspected to be near the quantum critical point on AFM long range order side in the presence of $J = 0.15$ meV with $J/D \sim 0.1$ (for the effective $S = 1$ scheme, $3J/D \sim 0.3$) [30]. In this case, Higgs modes like the longitudinal magnon in the low-energy spin excitation spectra are possibly observable in the inelastic neutron scattering or Raman spectroscopy [22, 31-33]. The presented magnetic results and the constructed magnetic phase diagram suggest that $Ba_2FeSi_2O_7$ is an important example of the $S = 2$ quasi-2D square lattice Heisenberg antiferromagnet with a strong easy-plane anisotropy, providing a suitable playground to test ideas of intriguing low dimensional quantum magnetism.

## ACKNOWLEDGMENTS

We thank C. D. Batista, H. Zhang, Jong-Hyeok Choi and Dae-Hyun Bang for useful discussions and Choongjae Won for crystal growth. This work was supported by the Max Planck POSTECH/Korea Research Initiative, Study for Nano Scale Optomaterials and Complex Phase Materials

(2016K1A4A4A01922028) and Grant No. 2020M3H4A2084418 through the National Research Foundation (NRF) funded by MSIP of Korea. The work at Oak Ridge National Laboratory was supported by the U.S. DOE, Office of Science, Basic Energy Sciences, Materials Sciences and Engineering Division, under contract No. DE-AC05-00OR22725. We acknowledge the support of the National Institute of Standards and Technology, U.S. Department of Commerce, in providing the neutron research facilities used in this work. The work at Rutgers University was supported by the U.S. DOE under Grant No. DOE: DE-FG02-07ER46382.

**Appendix A: X-ray diffraction data for crushed single crystals**

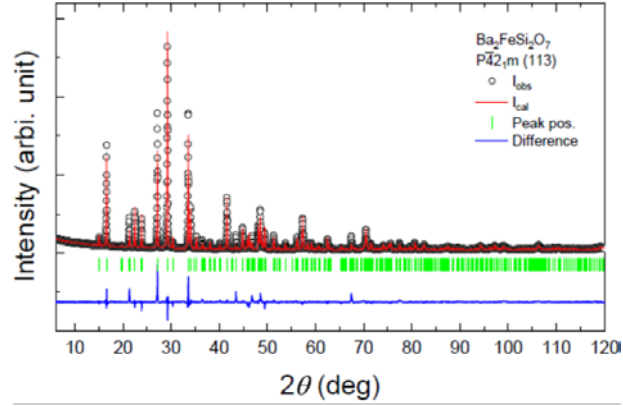

FIG. S1. X-ray powder diffraction pattern from crushed single crystals of $Ba_2FeSi_2O_7$ collected at $T = 300$ K. Open circles represent experimental data and solid line in red indicates a fitted line from Rietveld refinement using FULLPROF [26]. The blue solid line indicates the difference between experimental data and the fitted line. Green tick marker indicates the location of Bragg reflections for $Ba_2FeSi_2O_7$ phase. Crushed single crystals shows single phase of $Ba_2FeSi_2O_7$ but noticeable preferred crystallographic orientations are observed.

TABLE SI. Crystallographic information with space group $P\bar{4}2_1m$ (SG:113) from Rietveld refinements on the diffraction data at $T = 300$ K. Lattice constants $a = b = 8.3261(2)$ Å, $c = 5.3402(1)$ Å, and $\alpha = \beta = \gamma = 90°$. $R_{wp} = 20.2\%$.

| atom | *site* | x | y | z | B |
|---|---|---|---|---|---|
| Ba | 4e | 0.1693(1) | 0.6693(1) | 0.5098(4) | 1.33(2) |
| Fe | 2a | 0 | 0 | 0 | 0.60(9) |
| Si | 4e | 0.3689(5) | 0.8689(5) | 0.9665(13) | 1.17(16) |
| O1 | 2c | 0 | 0.5 | 0.1182(30) | 0.35(44) |
| O2 | 8f | 0.3452(15) | 0.8452(15) | 0.2767(19) | 1.85(34) |
| O3 | 4e | 0.0738(12) | 0.1996(11) | 0.1722(11) | 0.51(22) |

**REFERENCES**

* All the correspondence should be addressed to jhp@postech.ac.kr.

[1] N. D. Mermin and H. Wagner, Phys. Rev. Lett. **17**, 1133 (1966).

[2] A. Orendáčová, R. Tarasenko, T. Vladimír, E. Čižmár, M. Orendáč, and A. Fehe, Crystals **9**, 6 (2019).

[3] L. Onsager, Phys. Rev. **65**, 117 (1944).

[4] A. Cuccoli, T. Roscilde, V. Tognetti, R. Vaia, and P. Verrucchi, Phys. Rev. B **67**, 104414 (2003).

[5] Y. Kohama, M. Jaime, O. E. Ayala-Valenzuela, R. D. McDonald, E. D. Mun, J. F. Corbey, and J. L. Manson, Phys. Rev. B **84**, 184402 (2011).

[6] J. M. Kosterlitz and D. J. Thouless, J. Phys. C: Solid State Phys. **6**, 1181 (1973).

[7] A. Cuccoli, T. Roscilde, R. Vaia, and P. Verrucchi, Phys. Rev. B **68**, 060402 (2003).

[8] N. Tsyrulin, F. Xiao, A. Schneidewind, P. Link, H. M. Rønnow, J. Gavilano, C. P. Landee, M. M. Turnbull, and M. Kenzelmann, Phys. Rev. B **81**, 134409 (2010).

[9] P. Sengupta, C. D. Batista, R. D. McDonald, S. Cox, J. Singleton, L. Huang, T. P. Papageorgiou, O. Ignatchik, T. Herrmannsdörfer, J. L. Manson, J. A. Schlueter, K. A. Funk, and J. Wosnitza, Phys. Rev. B **79**, 060409 (2009).

[10] E. Čižmár, S. A. Zvyagin, R. Beyer, M. Uhlarz, M. Ozerov, Y. Skourski, J. L. Manson, J. A. Schlueter, and J. Wosnitza, Phys. Rev. B **81**, 064422 (2010).

[11] P. Sengupta, A. W. Sandvik, and R. R. P. Singh, Phys. Rev. B **68**, 094423 (2003).

[12] C. Yasuda, S. Todo, K. Hukushima, F. Alet, M. Keller, M. Troyer, and H. Takayama, Phys. Rev. Lett. **94**, 217201 (2005).

[13] H. Murakawa, Y. Onose, S. Miyahara, N. Furukawa, and Y. Tokura, Phys. Rev. Lett. **105**, 137202

(2010).

[14] I. Kézsmárki, N. Kida, H. Murakawa, S. Bordács, Y. Onose, and Y. Tokura, Phys. Rev. Lett. **106**, 057403 (2011).

[15] S. Bordács, I. Kézsmárki, D. Szaller, N. Kida, H. Murakawa, Y. Onose, R. Shimano, T. Rõõm, U. Nagel, S. Miyahara, N. Furukawa, and Y. Tokura, Nat. Phys. **8**, 734 (2012).

[16] K. Penc, J. Romhányi, T. Rõõm, U. Nagel, A. Antal, T. Fehér, A. Jánossy, H. Engelkamp, H. Murakawa, Y. Tokura, D. Szaller, S. Bordács, and I. Kézsmárki, Phys. Rev. Lett. **108**, 257203 (2012).

[17] M. Soda, L.-J. Chang, M. Matsumoto, V. O. Garlea, B. Roessli, J. S. White, H. Kawano-Furukawa, and T. Masuda, Phys. Rev. B **97**, 214437 (2018).

[18] P. Barone, K. Yamauchi, and S. Picozzi, Phys. Rev. B **92**, 014116 (2015).

[19] T. T. Mai, C. Svoboda, M. T. Warren, T.-H. Jang, J. Brangham, Y. H. Jeong, S.-W. Cheong, and R. Valdés Aguilar, Phys. Rev. B **94**, 224416 (2016).

[20] W. Low and M. Weger, Phys. Rev. **118**, 1119 (1960).

[21] D. Dai, H. Xiang, and M.-H. Whangbo, J. Comput. Chem. **29**, 2187 (2008).

[22] S.-H. Do, H. Zhang, T. J. Williams, T. Hong, V. O. Garlea, T.-H. Jang, S.-W. Cheong, J.-H. Park, C. D. Batista, and A. D. Christianson, "Decay and renormalization of a higgs amplitude mode in a quasi-two-dimensional antiferromagnet," (2020), arXiv:2012.05445 [cond-mat.str-el].

[23] B. E. Warren, Z. Krist. **74**, 131 (1930).

[24] V. Hutanu, A. Sazonov, H. Murakawa, Y. Tokura, B. Náfrádi, and D. Chernyshov, Phys. Rev. B **84**, 212101 (2011).

[25] M. Soda, M. Matsumoto, M. Månsson, S. Ohira-Kawamura, K. Nakajima, R. Shiina, and T. Masuda, Phys. Rev. Lett. **112**, 127205 (2014).

[26] J. RodrÃguez-Carvajal, Physica B: Condensed Matter **192**, 55 (1993).

[27] A. Wills, Physica B: Condensed Matter **276-278**, 680 (2000).

[28] C. Kittel, Introduction to Solid State Physics (J. Wiley, 1995).

[29] E. Gamsjager and M. Wiessner, Monats Chem **149**, 357 (2018).

[30] Z. Zhang, K. Wierschem, I. Yap, Y. Kato, C. D. Batista, and P. Sengupta, Phys. Rev. B **87**, 174405 (2013).

[31] M. Matsumoto, J. Phys. Soc. Jpn. **83**, 084704 (2014).

[32] A. Jain, M. Krautloher, J. Porras, G. H. Ryu, D. P. Chen, D. L. Abernathy, J. T. Park, A. Ivanov, J. Chaloupka, G. Khaliullin, B. Keimer, and B. J. Kim, Nat. Phys. **13**, 633 (2017).

[33] S.-M. Souliou, J. Chaloupka, G. Khaliullin, G. Ryu, A. Jain, B. J. Kim, M. Le Tacon, and B. Keimer, Phys. Rev. Lett. **119**, 067201 (2017).